**Paper presented as an original article, proposed for the Section "Physical Instruments for Ecology, Medicine and Biology"**

# A Simple Laser-Based Device for Simultaneous Microbial Culture and Absorbance Measurement

X. C. Abrevaya[a]; E. Cortón[b]; O. Areso[a], P. J. D. Mauas[a]

[a] Instituto de Astronomía y Física del Espacio (IAFE), UBA- CONICET. Ciudad Universitaria, Ciudad Autónoma de Buenos Aires (1428), Argentina.
[b] Laboratorio de Biosensores y Bioanálisis, Departamento de Química Biológica, Facultad de Ciencias Exactas y Naturales, UBA and IQUIBICEN-CONICET. Ciudad Universitaria, Ciudad Autónoma de Buenos Aires (1428), Argentina.

Corresponding author: Dr. X. C. Abrevaya.
Address: Instituto de Astronomía y Física del Espacio (IAFE), UBA- CONICET. Ciudad Universitaria, Ciudad Autónoma de Buenos Aires (1428), Argentina. Phone: (+54)-11-4789-0179, Ext.: 105. Fax: (+54)-11- 4786-8114. E-mail: abrevaya@iafe.uba.ar

**Abstract**

In this work we present a device specifically designed to study microbial growth with several applications related to environmental microbiology and other areas of research as astrobiology. The Automated Measuring and Cultivation device (AMC-d) enables semi-continuous absorbance measurements directly during cultivation. It can measure simultaneously up to 16 samples. Growth curves using low and fast growing microorganism were plotted, including: *Escherichia coli*, and *Haloferax volcanii*, an halophilic archaeon.

## INTRODUCTION

Microbial cell growth quantification is essential in different areas of research involving microbiological work. To implement it, several methodologies have been developed, which differ mainly in the physical or chemical properties employed to take the determinations [1].

Traditionally, growth curves are obtained manually, sampling culture aliquots, which are measured using a spectrophotometer or a similar optical instrument, usually obtaining a few experimental points that are used to plot a continuous curve, which could lead to errors in the determination of the growth parameters [2, 3]. In particular for microorganisms that grow slowly (generation time longer than 2 hours) this method could be very demanding, and to perform studies in this kind of microorganisms results unfeasible. In particular, it is not easy to detect this kind of microorganisms or their communities in environmental samples using conventional turbidimetry or plating methods.

A diversity of instruments were built to automatically measure the growth of different kind of microorganisms, not only on Earth but also in space [4-10]. Most are simple, inexpensive and automatic devices designed according to specific needs or applications [11-13].

Similar automated devices are commercially available, like the Turbidity Transmitter Trb8300® (Metller Toledo GmbH, Switzerland) which only allows instantaneous measurements. On the other hand, Bioscreen C® (Oy Growth Curves Ab Ltd, Finland) allows automatically obtaining growth curves, but it is an expensive instrument. Later Brewster [4] developed a device with very similar characteristics to Bioscreen C, but with lower cost, including the use of microwell plate technology.

In this paper we present an inexpensive alternative to Bioscreen C, the Automated Measuring and Culturing device (AMC-d), which enables continuous OD measurements by photometry on-line at real time, directly in the culture (*in situ*), running several samples simultaneously and consistently, which helps to perform duplicates or to test different conditions in parallel (multi-factorial experiments). The information is acquired, processed, and exported to a PC for the generation of microbiological growth curves, plotting turbidity vs. time, and the results are available remotely via the internet.

We demonstrate that the device works in a wide linear range where Lambert-Beer's law is applicable and that the measurements obtained using it has high reproducibility. As an example, we present growth curves for microorganisms from two different domains: the commonly used bacterium *Escherichia coli* (generation time around 20 minutes in optimum conditions) and *Haloferax volcanii*, an halophile archaeon which is an extremophile microorganism that grows at high salt concentrations (generation time around 3 hours in optimum conditions).

**DESIGN AN CALIBRATION**

The device basically consists of a structure built in acrylic which supports a rotating wheel made in aluminum with 16 slots where the samples can be placed using standard polymethyl methacrylate spectrophotometer cuvettes (1 cm optical path, total volume of 5 mL). A ring made of Delrin® is fixed on top of the rotating well using 4 stainless steel screws (Fig. 1). The wheel is moved (typically at 4.3 rpm) by a 12 Vcc motor, (model MR6-4.3, IGNIS, Buenos Aires, Argentina). This rotating wheel allows alternatively measuring each of the cuvettes, and mixing the cultures, providing aeration if an air chamber is left inside the cuvettes.

Attached to the structure, there are one 5 mW red laser diode module which emits at 655nm (M655-5, US Lasers Inc., CA, USA) and two silicon pin photodiodes (BPW34, Vishay Electronic GMBH, Germany) with a radiant sensitive area of 7.5 mm$^2$, one located behind the wheel (Ps, sample photodiode) and the other at 90º from the laser (Pr, reference photodiode). The laser beam is split by a semi-transparent mirrored glass: one beam is directed to Ps and the other to Pr. Ps measures the light from the laser after passing through the cuvette with the sample to be studied (or the corresponding blank/reference cuvette), and Pr measures the laser light passing through air, and is used to control the laser output (avoiding possible absorbance errors, originated in laser radiant power variations) (Fig. 1). The absorbance A is calculated as the logarithmic relationship between the radiant power passing through a cuvette containing a blank or reference solution (Po) and the one containing the sample (P), by using the light detected by Ps photodiode, as shown in Equation 1.

$$A = \log 10\ (Po\ P^{-1}) \qquad (1)$$

In some experiments, the concentration of a given molecule can be calculated by using the well know Lambert-Beer´s law (Equation 2), which establish that between a given concentration range, the absorbance is proportional to the molar absorbance of the analyte (ε), the optical path (b), and the concentration (c).

$$A = \varepsilon \, b \, c \qquad (2)$$

*Microcontroller and Programation*

External to the instrument, there is a command box containing an 8-bit AVR microcontroller, with 10-bit Analog to Digital Converter (Atmel, USA), an LCD Display (16 character x 2 lines converter ATmega16, Atmel, USA) and a numerical keyboard for operation. The command box is connected to a computer through an RS-232 port for data storage and to configure several parameters related to the measuring process and the shaking of the samples. These parameters are: the frequency of the measurements (in the 1-250 min range), the interval of time between two measurements, the timeout (which defines how much time is employed to stabilize the optical reading, in the 1-250 s range), the number of cuvettes to be measured (up to 15 samples plus a blank cuvette simultaneously) and the number of measurements (i.e. the length of the experiment, up to 3500 measurements). Automatic blank subtraction is possible, using a value registered in one cuvette, to be subtracted from the other values. Between measurements, the AMC-d works shaking the samples turning $N$ times with a time interval $t$, where both $N$ and $t$ are parameters that can be adjusted.

For each measurement, the date is registered (in format *day, hour* and *minute*). There is also a sensor which registers the temperature, and it is possible to set an alarm, which will sound if the temperature in the incubator goes outside a pre-specified range. All the operation of the AMC-d is controlled by software (programmed in Basic and compiled to the microcontroller language). The data is acquired and processed by the device and sent to the computer where it is stored in tabular form in a text file. The interface was programmed in Visual Basic. At the same time, a web interface allows to plot the results or to see them in tabular form both locally or *via* the internet, by means of Apache HTTP Server 2.2.

*Linearity*

The linearity predicted by Lambert-Beer´s law was checked with a calibration curve (absorbance vs. concentration) using a standard solution of $CuSO_4$. We prepared standard solutions of known concentrations (0.004 to 0.8 M), which were transferred to polymethyl methacrylate spectrophotometer cuvettes with lids that were placed into the AMC-d. The absorbance of each solution was measured and recorded. The same dilutions were also measured in a spectrophotometer (UVIKON 710, Kontron, USA) at a wavelength of 655nm. The calibration curves obtained were plotted and adjusted to a linear regression.

*Application in Microbiology Studies*

We assayed the ability of the AMC-d to obtain growth curves using two different microorganisms, with very different generation times: *Escherichia coli* (K-12 strain), a facultative anaerobic bacterium, and *Haloferax volcanii* (DS70 strain), an aerobic halophilic archaeon. *H. volcanii* DS70 strain was kindly provided by Dr. R.E. de Castro, Universidad Nacional de Mar del Plata, Argentina and *E. coli* K12 strain was obtained from ATCC.

*E. coli* was grown aerobically at 37 ºC until an OD (600 nm) =1.0 was reached. Tryptic Soy Broth (DIFCO) was used as culture media (30 g $L^{-1}$ in distilled water). *H. volcanii* was grown aerobically at 30 °C, to an OD (600 nm) =0.5. Growth medium Hv-YPC contains (g L$^{-1}$), yeast extract (5.0), peptone (1.0), casaminoacids (1.0), NaCl (144.0), MgSO$_4$.7H$_2$O (21.0), MgCl.6H$_2$O (18.0), KCl (4.2), CaCl$_2$ (0.35), and Tris-HCl (1.9); the pH was adjusted to 6.8.

The cultures for *E. coli* and *H. volcanii* were 1:10 diluted, and aliquots of 2.5 mL were placed in sterile standard polymethyl methacrylate spectrophotometer cuvettes, which were sealed and introduced into the AMC-d wheel. The device was placed inside an incubator, and the temperature was set at 37 ± 1ºC for *E. coli* and 30 ± 1ºC for *H. volcanii* (two consecutive experiments). The acquisition parameters were configured to the following values: one measurement every 5 min for *E. coli* and one every 60 minutes for *H. volcanii*, the timeout was 1 second, and between measurements the wheel turned 5 times every 30 seconds, to mix the samples. We also measured non-inoculated medium (blank cuvette) in the same conditions, to provide the value (Po) for absorbance calculation (Equation 1).

**RESULTS AND CONCLUSIONS**

We compare calibration curves using a colored ion (Cu$^{+2}$, from CuSO$_4$) as habitually used in UV-Vis spectrophotometer verification. The calibration data obtained with the AMC-d and a commercial spectrophotometer as a function of the CuSO$_4$ concentration show a very high linearity, with a correlation coefficient ($R^2$) of 0.999 (data not shown). When the errors were inspected, we found similar values, since the slopes were 2.865 ± 0.018 and 2.861 ± 0.009 for the AMC-d and the commercial spectrophotometer, respectively.

We also plot a regression line with the calibration data obtained from the two devices (data not shown), obtaining a correlation of 1.001 ± 0.006, indicating that our AMC-d behaves, indeed, linearly, between 0.1 and 2.3 and the values measured coincide very well with the ones obtained with the spectrophotometer within this range (to be compared with 0.008 to 1.200 for commercial devices as Bioscreen C) [3].

In order to study our device in systems where the light scatter produced by suspension particles is high, we measured different concentrations of an *E. coli* culture with the AMC-d and the spectrophotometer. The linearity of the AMC-d was indeed very good, with a correlation coefficient ($R^2$) of 0.998 (data not shown).

Then, *E. coli* and *H. volcanii* were cultivated using the AMC-d in two consecutive experiments. In particular, to study the reproducibility of the measurements, fourteen cuvettes with *E. coli* cultures were cultivated and measured using the AMC-d. To demonstrate how the AMC-d behaves for longer times of microbial growth, we cultivated and measured triplicate *H. volcanii* cultures for five days. For both experiments the OD values were obtained, through time, in the conditions mentioned in the previous section. The obtained data were averaged and plotted. In both cases the growth curves were obtained and the high reproducibility of the measurements can be observed (Fig. 2 and 3)

The AMC-d can be considered an inexpensive alternative to Bioscreen C and other similar devices, and also a prototype over which additional improvements could be made.

The main characteristics of the AMC-d are:

i- Growth is measured in real time, on-line, and automatically, providing a semi-continuous curve, where the individual measurements can be obtained with the desired frequency.

ii- Growth can be monitored directly in the culture (*in situ*) without risk of contamination, since there is no need to take culture aliquots to be measured in a conventional spectrophotometer.
iii- It is possible to measure several samples simultaneously and consistently, performing duplicates or testing different conditions in parallel (multi-factorial experiments).
iv- Although this device could be employed for any cultivable microorganism, it is particularly useful for microorganisms which grow slowly, or that cannot be measured with other methodologies (e.g: do not form isolated colonies, or take several weeks of incubation to form colonies).

Moreover, this device presents some advantages:

The components are economic and easy to find in international and local markets; they can easily be implemented in such way research laboratories can construct *ad-hoc* devices for their specific necessities [14-16]. The AMC-d presented in this work is one example of such equipment. Furthermore, the AMC-d can be easily assembled and disassembled replacing the pieces easily. The cuvettes can be discarded or re-used if an appropriate sterilization method is used. Moreover, due to its small size, low weight, low power consumption and simple design, the AMC-d is a portable device; it allows performing field studies.

The disadvantages of the AMC-d are related to the limitations of turbidimetry in general as a method, like interference caused by low homogeneity in the culture samples due to precipitates or, depending upon the microorganisms, of aggregate growing or flocculate formation and others associated with the limits of detection. It was determined that the minimum detectable absorbance value using conventional turbidimetry is 0.01 (which is equivalent in average, to $1 \times 10^7$ cells ml$^{-1}$ for bacteria). Below this value it is necessary to use other methodologies.

Another problem probably linked to variations of the temperature inside the cuvettes, is related to the OD stability at the beginning of the experiment, seen as perturbations in the growth curves, also seen in Bioscreen C [3, 17].

Also, the larger working volume allows recovering or treating the sample which is being measured, opening the possibility to carry out, after growing, another experiments or processes.

These characteristics makes AMC-d a useful device for environmental microbiology studies and in astrobiology since much research in this area involves the study of extremophile microorganisms and how their survival and growth is affected by different conditions expected to occur in hostile environments outside the Earth. Moreover, the fact that the results can be accessed remotely via the internet is convenient for these applications, and can also be used in laboratory experiments, giving the possibility to check the results remotely at any moment.

Even the prototype presented here works with a fixed wavelength (655nm), the laser source could be easily changed for another commercial one (wavelengths mostly used to measure bacterial optical density are 540, 600 and 660nm).

We also demonstrated the repeatability of the measurements, performing a large number of replications in our tests. Finally, the AMC-d is also stable and can be used in long experiments, as we showed in the *H. volcanii* experiment. The obtained results show that the AMC-d is a reliable instrument that allows the easy acquisition of growth curves in real time, either for slow or rapid growing microorganisms.

**ACKNOWLEDGEMENTS**

We would like to thank Guillermo Lemarchand and Leonardo Cantoni for their suggestions that helped to improve this work. We are grateful to computer technician Matías Pereira for their valuable work developing the software for the remote sensing and growth curve plots. We also want to thank to CONICET and ANPCyT for economic support.

**REFERENCES**

1. Madrid R.E., and Felice C., *Crit. Rev. Biotechnol.*, 2005, vol. 25, no. 3, p. 97.
2. Watson B.W., Gauci CL, Blache L., and O'Grady F.W., *Phys. Med. Biol.*, 1969, vol. 14, no. 4, p. 555.
3. Begot C., Desnier I., Daudin J.D., et al., *J. Microbiol. Meth.*, 1996, vol. 25, no. 3, p. 225.
4. Brewster, J.D., *J. Microbiol. Meth.*, 2003. vol. 53, no. 1, p. 77.
5. Wentink P., and La Rivière J.W., *Antonie Van Leeuwenhoek*, 1962, vol. 28, no. 1, p. 85-90.
6. Ševščik F., Liška B., and Hošek B., *Folia Microbiol. (Praha)*, 1963, vol. 9, no. 2, p. 125.
7. Platt T.B., Gentile J., and George M.J., *Ann. NY Acad. Sci.*, 1965, vol. 130, no. 2, p. 644.
8. Tsuji K., Griffith, D.A., and Sperry. C.C., *Appl. Microbiol.*, 1967, vol. 15, no. 1, p. 145.
9. Brown K.J., *Eur. J. Appl. Microbiol. Biotechnol.*, 1980, vol. 9, no. 1, p. 59.
10. Maxwell M., Allen E.R., and Freese E., *Appl. Environ. Microbiol.*, 1987, vol. 53, no. 3, p. 618.
11. Piccialli A., and Piscitelli S., *Rev. Sci. Instrum.*, 1973, vol. 44, no. 12, p. 1717.
12. Amrane A., and Prigent Y., *J. Microbiol. Meth.*, 1998, vol. 33, no. 1, p. 37.
13. van Benthem R., and de Grave W., *Microgravity Sci. Technol.*, 2008, vol. 21, no. 4, p. 349.
14. Ding J., Feng X., Song Y., and Wang W., *Instrum. Exp. Tech.*, 2011, vol. 54, no. 6, p. 849.
15. Sorouraddin M.-H., Rostami A., and Saadati M., *Food Chem.*, 2011, vol. 127, no. 1, p. 308.
16. Khokhlov A.A., Romanov R.A., Zubov B.V. et al., *Instrum. Exp. Tech.*, 2007, vol. 50, no. 3, p. 404.
17. Domínguez M.C., de la Rosa M., and Borobio M.V., *J. Antimicrob. Chemother.*, 2001, vol. 47, no. 4, p. 391.

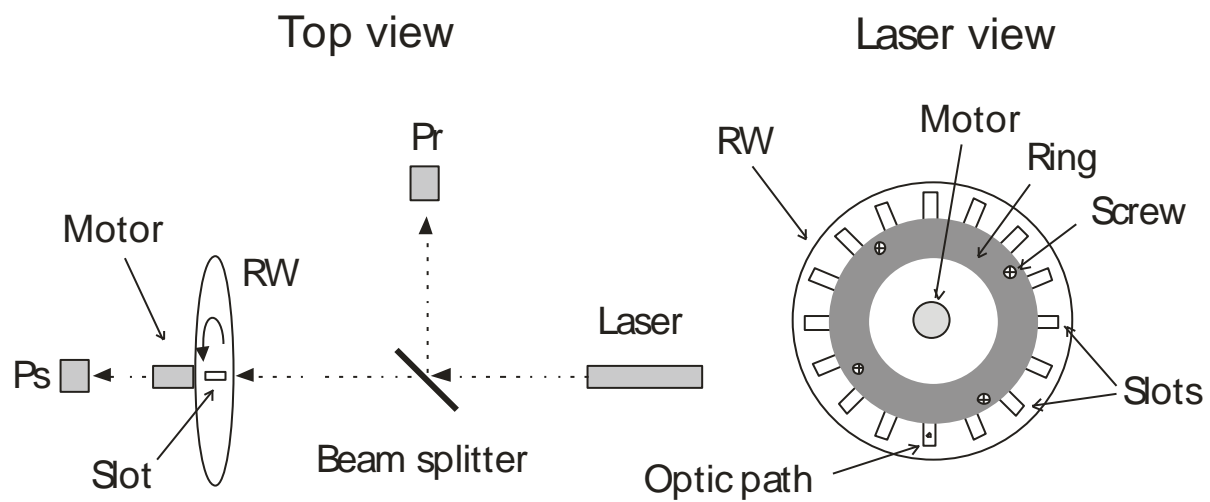

Fig. 1

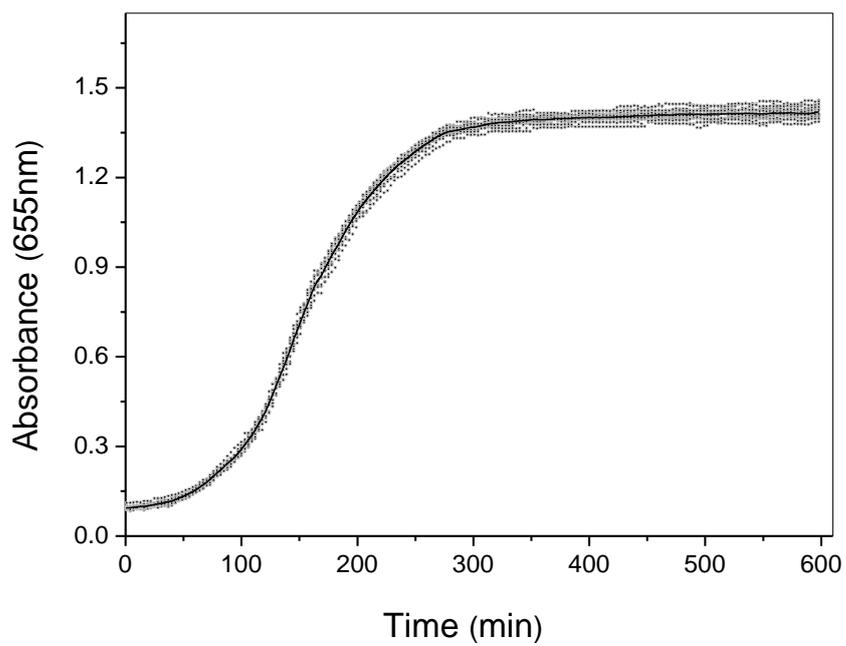

Fig. 2

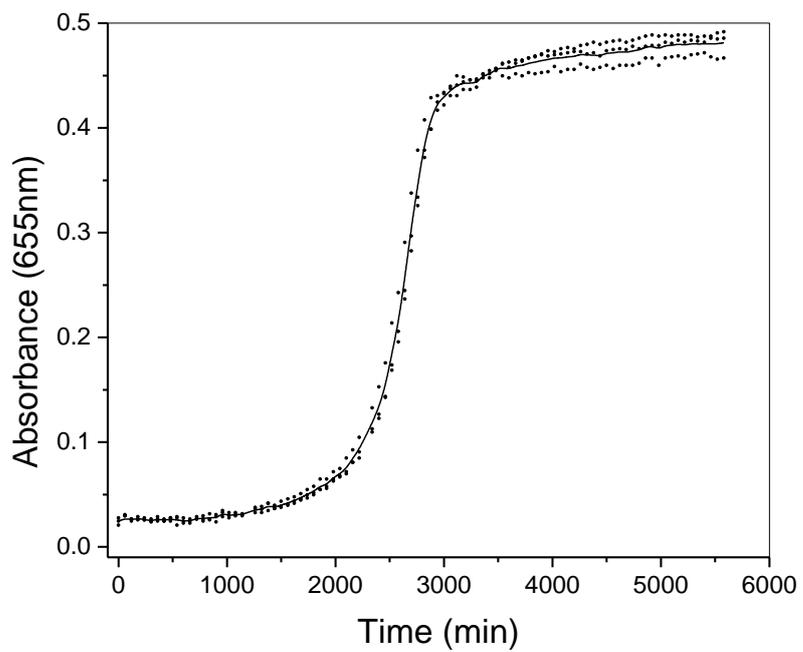

Fig. 3

**FIGURE AND TABLE CAPTIONS**

Fig.1. Schematic diagram of the device: (Ps) sample photodiode, (RW) rotating well, (Pr) reference photodiode.

Fig. 2. *E. coli* growth curve obtained using the AMC-d. The average of fourteen culture measurements is shown (continuous line). The residual values are also shown (dots).

Fig.3. *H. volcanii* growth curve obtained using the AMC-d. The average of triplicate culture measurements is plotted (continuous line). The residual values are also shown (dots).